# ON THE POSSIBLE EXPERIMENTAL DETECTION OF NON-CAUSAL SYNORDINATION PATTERNS OF PHYSICAL EVENTS

**Leonardo Chiatti**
AUSL VT Medical Physics Laboratory
Via Enrico Fermi 15, 01100 Viterbo (Italy)

**Summary**

With reference to a previous work, the problem of the experimental detection of non-causal synordination patterns between two series of physical events is examined. It is necessary that the patterns in question act in a *reproducible*, or at least *regular*, manner, so that they can be identified by means of statistical analysis; action on individual occasional facts cannot be proved statistically, though it is hypothetically possible.
As established in the previous work, the first series must be constituted by microevents (Penrose R processes), that can be amplified at the macroscopic level so that they are translated at this level into a choice between alternatives. The obtained result defines (by means of a structural constraint which is not, however, a causal link) the choice between execution or non-execution of certain actions on a macroscopic system S, without any feedback on the microevents source. A succession of macroevents is thus generated, which makes up the second series.
Statistics of the events belonging to the second series is thus completely defined by the probability distribution of the first series of events. In the absence of a synordination, this is the *a priori* probability distribution, which can be measured by removing the constraint between the two series. If there is synordination, the probability distribution of the two sets of events will deviate in a measurable manner from the expected trend once the constraint is activated.

**Introduction**

In a previous work (1) the possibility was discussed of acausal links between series of events in the physical world. We refer to that work for a detailed discussion of the mechanism on which these hypothetical links are based; our concern here is to formulate an experimental setup capable of proving its possible existence. Indeed, while synordination patterns corresponding to the causal links between events certainly exist, it remains to be proved that nature has availed itself of the possibility constituted by the more general acausal synordination.
First of all, it is necessary to summarize what we are seeking. Our attention is turned to microevents in which the wavefunction of a microparticle undergoes a reduction; for example (but not necessarily), microevents consisting of the preparation of the "initial state" or of the detection of the "final state" of the microparticle. Following a suggestion by Penrose (2), we shall call these events "R processes", where "R" stands for "reduction", in order to distinguish them for the following, or previous, unitary time evolution of the wavefunction ("U processes", where U stands for "unitary"). Though here we are using the wave formulation of quantum mechanics, the difference between U and R processes is universal and independent of the specific formulation adopted. For example, in the formulation based on the trajectories of the microparticle in its phase space, U processes are associated with non-Kolmogorovian joint probabilities, while R processes are always associated with marginal Kolmogorovian probabilities.

As a second condition, we require that certain groups or series of the microevents under consideration activate chains of events which culminate in a macroscopic effect; here we understand "macroscopic" as "describable according to the language and laws of classical, non-quantum and non-relativistic physics". The detection of the state of the microparticle by a measurement apparatus is an evident, though not exclusive, example of an amplification process of this type.

The third condition is that a certain, automatically established relation exists between each group or series of microevents and the macroscopic effect produced by it. In other words, the macroscopic effect must be a reading or translation of the group or series of microevents.

Let us now consider the set of all possible macroscopic effects induced by the groups or series of microevents taken into consideration. Using a pseudorandom number generator, let us assign each of these macroevents a quality A or (in the exclusive sense, as in the Latin *aut*) non-A. The A or non-A qualities must satisfy the law of the excluded middle and that of non-contradiction, in the sense that if a given macroevent is not A, then it is non-A and vice versa, and no third possibility (neither A nor non-A) is admitted.

The fourth condition is that if a type-A macroevent actually takes place, an effect is produced, in an automatic and necessary manner, on a certain macroscopic system S. If, instead, the macroevent which takes place is of the non-A type, the effect on S is not produced. *The total absence is assumed of causal retroactions of the macrosystem S on the microevent source.*

We propose to consider the existence of a synordination among the groups or series of microevents and the succession of macroevents that constitute the choices : action on S, non-action on S. If such a synordination exists, it must be acausal, since, according to our hypothesis, any causal retroaction of the macrosystem S on the source of microevents is excluded.

Regardless of whether or not a synordination exists, the probability distribution of the choices (A, non-A) is entirely determined by the probability distribution of the groups or series of microevents. If the totally automatic experimental apparatus is inhibited to convert the choice (A, non-A) into the choice (action on S, non-action on S) by removing the necessary triggering device, the probability distribution of the results (A, non-A) should be as expected according to the usual physical laws and will be called *a priori* probability.

When the trigger is inserted, the probability distribution of events (A, non-A) can remain the same or change with respect to the previous case. In this second case, we have an acausal synordination of the choices (action on S, non-action on S) and the groups or series of microevents.

The choice of a suitable macrosystem S remains highly based on conjecture. In ref. (1), the hypothesis has been made that the non-causal synordination can play a role in the ontogenesis and phylogenesis of biological systems. It is intriguing to think in these terms of the discontinuities of the phylogenetic process, widely documented in paleontology, as an alternative or in addition to more traditional theories such as punctuated equilibria. It seems therefore natural to select a biological system as S macrosystem. Action on S will favour (or, on the contrary, will hinder) the survival, growth or adaptation of the system, while non-action on S will have an opposite or neutral effect.

Even more generally, S could be, rather than a biological system, any system capable of self-evolution, and in this case the action on S could imply choosing the direction of an *irreversible* process within the context of this self-evolution. The S system could thus be any system capable of developing more sophisticated structural and functional orders through the use of energy and matter taken from the environment. With systems capable only of mechanical replication or conservation of their own internal order, synordination might not occur or in any case would have little effect.

# A possible experimental setup

Let us consider a beam of light coming from a source, even a very distant one, e.g. from the Sun. Let us extract from it, using strong attenuation and filtering, a few photons at a time, all identical. These identical photons are then sent to a beamsplitter, with a 50/50% probability of being transmitted/reflected.

Let us indicate as '1' the event of the transmission of an individual photon followed by the detection of the same photon by a photodiode. Let us indicate as '0' the event of the reflection of an individual photon followed by the detection of the same photon by another photodiode. A series of $n$ successive events of this type will thus correspond to an integer binary number with $n$ digits such as 00100111010101 or similar ones. The set of numbers that can be obtained in this way is the set of the first $2^n$ integer numbers.

Let us consider $2^n$ closed opaque boxes, in a fraction $p$ of which there is a just germinated plant. A progressive integer number starting from 0 will be associated with each box. The boxes containing the plants will be distributed in random order with respect to the progressive number assigned to each.

The occurrence of a series of $n$ microevents will automatically command the opening of the box associated with the corresponding progressive number for a period of 5 minutes. The inside of the box will be exposed to the Sun and, if there is a plant inside, it will enjoy a healthy 5 minutes of photosynthesis. After 5 minutes, the box will close again automatically and the experiment will be repeated.

By repeating the procedure $N$ times, with $N$ a very large number, the event "there is a plant in the open box" should take place on average $Np$ times, with a standard deviation of $[Np(1 - p)]^{1/2}$. A statistically significant[1] deviation from this prediction would imply the existence of an acausal link between two choices:

    transmitted or reflected photon, subsequently detected

    illuminated or non-illuminated plant.

The connection between these two choices, constituted by the random pre-selection of the boxes in which the plants are to be placed, does not constitute a causal link between them; it is merely a constraint. A certain succession of microevents (transmissions and reflections and successive detections) brings about the opening of certain boxes, as a result of the classical amplification of microscopic fluctuations (operation of light detectors). This is not in relation with the possible presence of plants inside these boxes.

As far as the individual microevent is concerned, the probability of the results 1 and 0 is still ½, even in the presence of synordination. Indeed, the plants are randomly distributed among the various boxes, and therefore the ratio between the average number of 0 and 1 figures in the binary codes of the boxes containing the plants is equal to that found for all the boxes in general. Thus there is no violation of the postulates of quantum mechanics.

Moreover, the events of transmission/reflection of distinct photons remain statistically independent trials, so that not even at this level are quantum laws violated. Rather, synordination would imply the appearance of *successions* of *independent* events with greater frequency than expected, which would favour a certain result.

A control run without plants can be performed, to check that there are no systematic errors due to the apparatus (e.g. photodiodes not all perfectly equally efficient). Different runs can then be performed with different box codes, in order to eliminate any artefact caused by their codification. The permutation associated with the recodification will have to be performed automatically by means of a relay system guided by a random or pseudorandom number generator, in order to exclude any human intervention which could account for any possible "psychokinetic" effects.
In this example, the source of microevents is the Sun + beamsplitter + photodiodes system; the characteristic A, non-A associated with each result, i.e. with each box, is given by the presence or absence, respectively, of a plant in that box. The system S is the set of plants.

---

[1] If $x$ is the number of favourable occurrences, the expected distribution of the variable $(x - Np)/[Np(1 - p)]^{1/2}$ tends, for $N \to \infty$, to the standard normal (0;1).

**Remarks**

1) If $p$ is the *a priori* probability that the event of the second series "the chosen box contains a plant" (calculated by applying only causal physical laws) and $p'$ is its probability actually measured as the frequency of its occurrence, let us consider the information surplus:

$$H = \log_2 (p'/p) \ .$$

It is not associated with any energy cost nor with any thermodynamic entropy variation of the detector-amplifier-trigger apparatus: when a box is opened, this apparatus exchanges the same energy and entropy whether the box contains a plant or not. The synordination can be detected if $H$ is not nil. Of course, one can well imagine acausal links consisting of the appearance of *individual* "providential" events, without any shift in the probability distributions involved by their expected values. In this case, however, we would have $p = p'$ and the acausal link would be experimentally indistinguishable from a fortuitous coincidence; these "providential" links can therefore not be identified using statistical methods.
The appearance of a share of information not associated with any energy or entropy exchange is a certain sign of a *creative* activity of the process (1). One must note, however, that energy and entropy exchanges of the detector-amplifier-trigger system and of the S system are always governed by the usual laws of thermodynamics, *also in the case of synordination*. In particular, the possible result $H \neq 0$ does not conflict with the II principle of thermodynamics.

2) To understand better the role of a possible acausal synordination between natural selection and random genetic mutations[2], the following modification of the experiment can be considered.
Let us cultivate a bacterial population that is as homogeneous as possible and let us isolate from it a mutant strain resistant to a given antibiotic X (we shall suppose that the mutation has been spontaneous and not induced by exposure to some mutagenic agent). Let us prepare a certain number of separate cultures, having equally numerous populations, both of the wild strain and of the mutant one. One of these cultures is placed in each of the $2^n$ boxes of the previous experiment. Cultures of the mutant strain are placed in a percentage $p$ of the boxes; cultures of the wild strain are placed in the remaining ones. The choice of the boxes in which the mutant strain cultures are to be placed is made using a random number generator.
The occurrence of a given series of $n$ microevents will correspond, as before, to the choice of a given box. An entirely automatic system will pour the antibiotic X into the culture inside the chosen box. In this version of the experiment, $p$ is the *a priori* probability that a box containing a culture of the strain resistant to the antibiotic X is chosen, while $p'$ is the effectively measured probability of the same event.
If $p'$ is proven to be different from $p$ in a statistically significant manner, then we have a synordination between the two series of events:

transmitted or reflected photon, subsequently detected

---
[2] An idea originally formulated by Pauli (5).

administering of the antibiotic to the resistant strain or to the wild one.

The control is carried out by preventing that the automatic device can expose the cultures to the antibiotic. The probability that a box with a culture of the mutant strain is selected in such "blank shot" must therefore coincide with $p$.

3) Obviously, in the real biosphere the detector-amplifier-trigger apparatus is internal to the biological system S and not external to it. It follows that the initial clause on the absence of retroaction of the system S on the apparatus certainly ceases to apply. This makes it extremely difficult to identify non-causal synordination (if it exists) when directly observing biological phenomena. If the difference $p'- p$ implies a retroaction on the quantum level (equivalent, for example, to increasing the transmissivity of the beamsplitter at the expense of its reflectivity or vice versa), the effect is to introduce the system S into an attractor (metaphorically: in the end the plants will all be endlessly exposed, all endlessly obscured or alternately exposed to light and darkness in a limit cycle).

4) Furthermore, in real situations we shall not only have successions of microevents 0 and 1 over time but also their coincidences within a certain time interval. It is possible that logical choices made on these groups of simultaneous events by means of Boolean connectors "and", "or", "not", imply, as a macroscopic effect, the discharge/non-discharge of previously excited devices (e.g. condensers). These devices are the equivalent of our "boxes" and closely resemble the neurons in the central nervous system of a vertebrate.

5) It is important to emphasise that in the proposed experiments only physical phenomena are involved and in no case psychic or mental phenomena, such as intentional choice, for example. If the possibility of a *voluntary* choice were introduced, one could decide, for example, whether or not to open the box with the plant *after* the occurrence of a macroevent A. Consequently, the exposure of the plant to sunlight would no longer be an automatic and necessary effect of the macroevent. The only way to avoid a temporal paradox taking place would thus be to suppose that at that point $p'$ changes, becoming once again equal to $p$. This would lead to the disappearance of the synordination.

**Goethean archetypes vs synordering**

In ref. (1) the notion of "synordination pattern" is associated with that of "archetype" while the "similitude metamorphosis" is compared with the Goethian concept of metamorphosis. Recently, some Goethian authors have drawn attention to the opportunity of what is known as "delicate empiricism", an epistomological approach based on intuition, through which direct perception of archetypes (3,4) would be attained. Clearly, by accepting the proposal of these authors in its extreme form, the entire experimental research program outlined in this article appears to be superfluous, if not wrong, as the mediation of experimental devices is proposed in order to detect a reality which ought instead to be directly accessible through intuitive experience.
The point is that we are actually talking about two different things and calling them by the same name. Let us suppose that two different sets of microevents, capable of being amplified up to the macroscopic level and thus giving rise to macroprocesses compatible with two different explicate orders $E$ and $E'$, are connected through a similitude metamorphosis $E' = MEM^{-1}$. This metamorphosis will have a structure, which will be part of a certain more general synordination pattern which we call *archetype*. In this meaning, the archetype is therefore a *process* pattern, which connects the explicate order to itself through the implicate order. It only acts indirectly on the world of macroscopic *forms*, because it actually concerns microprocesses, which cannot be observed with the naked eye and probably are inaccessible to intuition. These structures can only be found with an objective experimental investigation, such as that proposed here.

On the other hand, the explicate orders *E* and *E'* are distinguished by various macroscopic features, and according to some of these features they can be placed at the vertices of a tree diagram representing a *taxonomy*. For example, *E* and *E'* could be distinguished by the colour of the leaves. Thus on an appropriate taxonomic tree there will be a vertex associated with "having green and red leaves", with two branches, one of which will lead to vertex E (green leaves) and the other to vertex E' (red leaves).

In any case, one might wish to distinguish *E* from *E'* not by the colour of the leaves but by their having long or short thorns. On another suitable taxonomic tree there will thus be a vertex "with thorns", with two branches leading respectively to *E* (short thorns) and *E'* (long thorns).

Now, these classifications are entirely conventional logical constructions and not objective real processes as synordering is supposed to be. Furthermore, we are discussing here static relations between equally static *forms* and not processes that effectively connect one form with another, either within or outside of time. Taxonomic trees are conventional mental structures representative of a descriptive order for which there is no corresponding generative order, which instead constitutes the essential aspect of synordering. On this basis, it is certainly true that the homologies and analogies between forms can be perceived through intuition, but this has actually nothing to do with the issues discussed in this article. From our point of view, the Ebach homologies are an *expression* of a process-archetype, not the *image* of a form-archetype.